\documentclass[conference]{IEEEtran}
\usepackage[cmex10]{amsmath}
\usepackage{graphicx}
\usepackage{cite}
\usepackage{epstopdf}
\usepackage{algorithm}
\usepackage{algorithmic}
\usepackage[caption = false]{subfig}
\usepackage{booktabs}
\usepackage{amsfonts}
\usepackage[letterpaper, left=0.65in, right=0.65in, bottom=1in, top=0.8in]{geometry}

\begin{document}
\title{Energy-Efficient User Access Control and Resource Allocation in HCNs with \mbox{Non-Ideal} Circuitry}
\author{\IEEEauthorblockN{Yuhao~Zhang, Qimei~Cui, and Ning Wang}
\IEEEauthorblockA{School of Information and Communication Engineering,\\
Beijing University of Posts and Telecommunications, Beijing, 100876, China\\
Email: cuiqimei@bupt.edu.cn}}
\maketitle

\begin{abstract}
In this paper, we study the energy-efficient user access control~(UAC) based on resource allocation~(RA) in heterogeneous cellular networks~(HCNs) with the required downlink data rate under \mbox{non-ideal} power amplifiers~(PAs) and circuit power. It is proved that the energy consumption minimization is achieved when the typical user accesses only one base station~(BS), while the other BSs remain in idle mode on the transmission resource allocated to this user. For this purpose, we reformulate the original \mbox{non-convex} optimization problem into a series of convex optimization problems where, in each case, the transmit power and duration of the accessed BS are determined. Then, the BS with the minimal energy consumption is selected for transmission. Considering the approximate situation, it is showed that the optimal transmit duration of the accessed BS can be estimated in closed form. The benefits of our proposed UAC and RA schemes are validated using numerical simulations, which also characterize the effect that \mbox{non-ideal} PAs have on the total energy consumption of different transmission schemes.
\end{abstract}

\vspace{2mm}


\section{Introduction}\label{sec:1}
Experiencing the explosive data growth over the recently years, the mobile wireless communications have developed swiftly and require larger system capacity, stronger signal coverage, and higher energy efficiency~(EE) than the past~\cite{Refenece1}. With these developments, the wireless networks in the future need the user-centric framework with the improved quality of service~(QoS) for all kinds of users, and have to transmit numerous data information with higher spectral efficiency~(SE)~\cite{Refenece21}. Therefore, in order to solve these problems, the architecture of heterogeneous cellular networks~(HCNs) has been proposed as a promising technology to increase the system capacity, strengthen the coverage performance and improve the user experience~\cite{Refenece5,Refenece6}. However, since HCNs need to deploy more hardware infrastructures, which inevitably cause rapidly grown energy consumption, green communication for HCNs has drawn much attention all over the world~\cite{Refenece8}. Therefore, the topic of energy-efficient HCNs is very interesting and worth to research, which is mainly discussed in this paper.

In HCNs, since numerous transmission nodes are deployed irregularly, there may be some areas that are covered by more than one BS, in which users can have a better chance to access the networks. In particular, due of the maximal received useful signal power, the optimal SE for the typical user can be achieved by accessing as many BSs as possible, where the user can be served by all the accessed BSs in cooperative transmission~\cite{Refenece10,Refenece22}. Therefore, with required data rate, it is intuitive that accessing more BSs can save the energy consumption on account of the reduced transmit power. However, in practice, considering both the non-negligible circuit power and the fact that a great part of the power input to a \mbox{non-ideal} power amplifier~(PA) are not used for data transmission, this method may lead to higher overall energy consumption of HCNs due to the activation of all BSs, even if the transmit power can reduced by cooperative transmission~\cite{Refenece22,Refenece11}. As a result, the user access control~(UAC) for minimal energy consumption under \mbox{non-ideal} circuitry is quite meaningful, but as far as we know, there is no certain result in the existing works.

The UAC, associating a user with its serving BSs, is of great importance to the system performance of HCNs, and has been widely studied in the existing literatures~\cite{Refenece12}. Traditionally, the UAC is fulfilled based on the maximum received signal strength provided by the single serving BS~\cite{Refenece13}, while more sophisticated UAC schemes are proposed in the emerging wireless networks for different goals, e.g., signal analysis, QoS guarantee, performance trade-off, SE and EE~\cite{Refenece14,Refenece20}. However, the \mbox{non-ideal} circuitry, which has a significant effect on UAC in HCNs, has not been considered in all the relevant literatures. In practice, PA responses are \mbox{non-ideal}, where the power efficiency varies \mbox{non-linearly} with respect to the output power~\cite{Refenece18,Refenece19}. The presence of \mbox{non-ideal} PAs creates a \mbox{non-convex} structure in the objective function, and the optimization for transmit duration is introduced due to circuit power, which both complicate the optimization problem.

In this paper, given the required data rate, considering two different types of channel state information~(CSI), we study the energy-efficient UAC by investigating the resource allocation~(RA) with minimal total energy consumption in HCNs under \mbox{non-ideal} PAs and \mbox{non-negligible} circuit power. For this purpose, we first formulate the \mbox{non-convex} energy optimization problem with respect to transmit powers and duration. It is proved that unlike the SE maximization, the minimal overall energy consumption is obtained when only one BS is accessed by the typical user and other BSs maintain silence on the same resource, based on which the original \mbox{non-convex} problem is transformed into a series of convex optimization problems, where the optimal transmit duration can be resolved by using linear search. Moreover, in the approximate situation, a closed form solution for the transmit duration can be obtained. Simulations are carried out to validate the benefits of our proposed scheme, and characterize the effect of \mbox{non-ideal} circuitry on the energy consumption.

The rest of the paper is organized as follows. Section~\ref{sec:2} described the system model and formulated energy minimization problem, which was resolved in Section~\ref{sec:3}, followed by algorithm in Section~\ref{sec:4}. Simulation results were provided in Section~\ref{sec:5}, followed by conclusion in Section~\ref{sec:6}.
\section{System Model and Problem Formulation}\label{sec:2}
\begin{figure}[!t]
\centering
\vspace{2mm}
\includegraphics[scale=0.38]{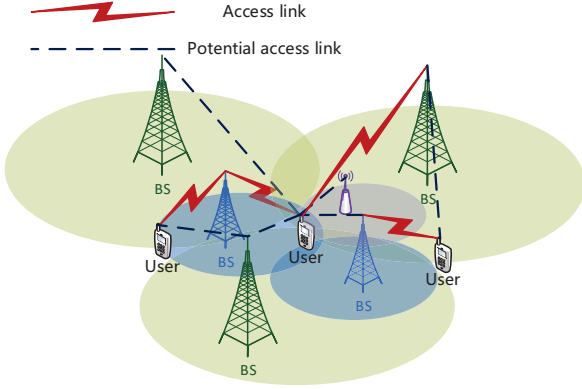}
\caption{System model for UAC in HCNs.}
\label{System_model}
\vspace{-4mm}
\end{figure}

In the considered HCNs, the BSs are deployed in a researched area, conforming a two-dimensional spatial homogeneous Poisson point process~(PPP). It is assumed that a typical user equipment~(UE) located at the origin $(0,0) \in \mathbb{R}^2$ lies in the intersection of signal coverage of $M$ BSs. The typical UE can access a subset of the $M$ BSs, denoted by $\mathcal{M}$, which cooperatively transmit the same data information, using the same transmission resource, during every scheduling time interval of duration~$T$, as shown in Fig.~\ref{System_model}. \mbox{High-speed} \mbox{low-latency} backhaul links, e.g., optical fibers, and perfect \mbox{time-frequency} synchronization are assumed among $\mathcal{M}$. The CSI for all BSs is available and can be exchanged through backhaul links without any error and delay. A block fading channel, including large-scale and small-scale fadings, is used to characterize the complex channel gain, denoted by $h_m, m \in \mathcal{M}$, from BS~$m$. The system bandwidth is $W$, and $n$ is additive white Gaussian noise~(AWGN) at the UE with power $P_N=N_0 W$, where $N_0$ is the power spectral density~(PSD) of the noise.
\subsection{Signal Model}\label{sec:2a}
In the scheduling time interval, the cluster $\mathcal{M}$ transmits data jointly to the UE within $t \leq T$ seconds and turn into sleep mode during the rest of the frame, i.e., $(T-t)$ seconds. Therefore, the received signal at the typical UE during the active time can be formulated as
\begin{equation}\label{ReceSignal}
    y=\sum_{m \in \mathcal{M}} \sqrt{p_{m}} \, {h}_{m} \, x_m + i_{\rm out} + n,
\end{equation}
where $p_m$ and $x_m = (w_m \, s)$ are the transmit power and copy of the common information symbol $s$ that BS~$m$ transmits, respectively. Note that $w_m$ is the weighting factor at BS $m$ and $\mathbb{E}\big\{ {{{\left| {{x_m}} \right|}^2}} \big\} = 1$. And $i_{\rm out}$ is overall interference with power $I_{\rm out}$, which is created by other BSs outside $\mathcal{M}$. Due to numerous sources and randomness of the interference, it is modeled by the AWGN for average performance.

In this paper, there are two methods of CSI feedback, i.e., the \mbox{short-term} and \mbox{long-term} form. In \mbox{short-term} form, both the amplitude and phase information $\{{| {{h_m}} |}, e^{j \angle{h_m}}\}$ of the channel can be sent back, however in \mbox{long-term} form, only the amplitude information $\{{| {{h_m}} |}\}$ is available. Depending on the type of CSI, there are two kinds of weighting factor $w_m$ considered in this paper. With \mbox{long-term} CSI, the cluster $\mathcal{M}$ only can transmit at the same time directly due to the unknown phase information, while with \mbox{short-term} CSI, phase compensation is made first before transmission to enhance the received signal power. Therefore, the received signal power in the two cases can be expressed respectively, as given by
\begin{equation}\label{Useful_Power}
S  =
\left\{ \begin{array}{ll}
\sum\limits_{m \in \mathcal{M}} {{p_m}{{\left| {{h_m}} \right|}^2}}, & {\text{\mbox{long-term} CSI}},\\
{\left( {\sum\limits_{m \in \mathcal{M}} {\sqrt {{p_m}} {\left| {{h_m}} \right|}} } \right)}^2, & {\text{\mbox{short-term} CSI}}.
\end{array} \right.
\end{equation}

Finally, according to Shannon formula, the achievable data rate attains the form, as presented by
\begin{equation}\label{AchiRate}
    C = \left(\frac{t}{T}\right) \, W \, \log_2 \left( 1 + \frac{S}{I_{\rm out} + P_{\rm N}} \right).
\end{equation}
\subsection{Energy Consumption Model}
\label{sec:2b}
The efficiency $\eta$ of a PA is defined as the ratio between the output transmit power and the actual total power consumption. Moreover, the traditional power amplifier~(TPA) is considered in this paper~\cite{Refenece18,Refenece19}, where the total power consumption for BS $m$ can be given by
\begin{equation}\label{TPA}
\Psi_{\rm tpa}(p_m) = \frac{\sqrt{p_m P_{\max,m} }}{\eta_{\max,m}}, \quad m \in \mathcal{M},
\end{equation}
where $P_{\max,m}$ and $\eta_{\max,m}$ are the transmit power parameter and maximum PA efficiency in BS~$m$, respectively.

Considering TPA model and circuit power, the total power consumption for transmission of BS $m$ can be written as
\begin{equation}\label{ptx}
    P_{{\rm tx},m}= \Psi(p_m) + \varepsilon_{m} r_{\rm dl} + P_{{\rm base},m},
\end{equation}
where the circuit power consumption is divided into two parts, i.e., static and dynamic components. The static component $P_{{\rm base},m}$ is the constant, whereas the dynamic component $p_{{\rm c},m} = \varepsilon_m \, r_{\rm dl}$, which depends on both the actual downlink data rate $r_{\rm dl}$ and power consumption coefficient for transmitting a data rate unit, denoted by $\varepsilon_m$.

Similarly, the power consumption for reception at the UE can be given by
\begin{equation}\label{prx}
    P_{{\rm rx},{\rm u}}=\varepsilon_{\rm u} \, r_{\rm dl} + P_{\rm{base},{\rm u}}.
\end{equation}

Finally, the power consumption in idle mode is assumed to be the constant (i.e., independent of $r_{\rm dl}$) at both BS $m$, denoted by $P_{{\rm idle},m}$, and the typical UE, denoted by $P_{\rm{idle},{\rm u}}$.

\subsection{Optimization Problem Formulation}
In this subsection, we consider the energy consumption minimization in HCNs with required downlink data rate ${r_{\rm dl}}$ under \mbox{non-ideal} PAs and \mbox{non-negligible} circuit power. In this case, the energy minimization problem can be formulated as
\begin{equation}\label{G_EE_Opt}\tag{\textbf{P1}}
\begin{aligned}
    \mathop {\min }_{\{p_m\},t} \quad & \sum_{m = 1}^M  \mathbb{I}(p_m) \Big[ {P_{{\rm tx},m}} \, t + {P_{{\rm idle},m}} \, \big( T - t \big) \Big]  \\
    &+ \sum_{m = 1}^M \Big[ 1 - \mathbb{I}(p_m) \Big] {P_{{\rm idle},m}} T \\
    &+ \Big[ {P_{{\rm rx}, {\rm u}}} \, t + {P_{{\rm idle}, {\rm u}}} \big( T - t \big) \Big]
\end{aligned}
\end{equation}
\begin{equation}\nonumber
{\rm s.t.}\ \ {r_{\rm dl}} \le C,
\end{equation}
\begin{equation}\nonumber
0 \le t \le T,
\end{equation}
\begin{equation}\nonumber
p_m \geq 0,\ m = 1, \dots, M.
\end{equation}
where $\mathbb{I}(\cdot)$ is an indicator function, i.e., $\mathbb{I}(x) = 1$ if $x > 0$; otherwise, $\mathbb{I}(x) = 0$. Note that $P_m = 0$ if BS $m$ is not accessed. It is seen that for the optimal solution of~\textbf{(P1)}, ${r_{\rm dl}} = C$ must be satisfied, which can be easily proved by reducing transmit powers $\{p_m\}$ until the data rate equality holds. Note that for simplicity, it is assumed that transmit power $p_m$ will not exceed its maximum $P_{\max,m}$; in other words, $P_{\max,m}$ is large enough to support the required data rate $r_{\rm dl}$.

\section{Energy Minimization under TPA}\label{sec:3}
In this section, we will resolve the energy minimization problem~\eqref{G_EE_Opt} under TPA. Substituting \mbox{\eqref{TPA}--\eqref{prx}} into~\eqref{G_EE_Opt} and suppressing the constant term, the objective function can be transformed, as written by
\begin{equation}\label{EE_Opt_TPA}
\begin{aligned}
    E_{\rm tpa} ( \{p_m\}, t) &= \sum_{m=1}^M \bigg( \frac{\sqrt {{P_{\max ,m}} }}{{{\eta _{\max ,m}}}} \bigg) \,\sqrt{{p_m}} \, t \\
	& +  \sum_{m = 1}^M \mathbb{I}(p_m) \big( \varepsilon_m \, r_{\rm dl} + P_{{\rm base},m} - P_{{\rm idle},m} \big) \, t \\
    & +  \big( \varepsilon_{\rm u} \, r_{\rm dl} + P_{{\rm base},{\rm u}} - P_{{\rm idle},{\rm u}} \big) \, t.
\end{aligned}
\end{equation}
It is clear that the optimization problem (\ref{G_EE_Opt}) is not convex w.r.t. the transmit powers and duration, due to the \mbox{non-convexity} of objective function~\eqref{EE_Opt_TPA} and even the non-convex feasible region caused by the logarithmic data rate constraint. This characteristic can be confirmed by analyzing the corresponding Hessian matrix, as calculated by
\begin{equation}\nonumber
\left(
  \begin{array}{ccccc}
    -\frac{1}{4} \Gamma(p_1)^{-\frac{3}{2}} t \hspace{-3mm} \hspace{-1mm} & \hspace{-1mm} 0 \hspace{-0.5mm} & \cdots \hspace{-1mm} & \hspace{-1mm} 0 \hspace{-1mm} & \hspace{-1mm} \frac{1}{2} \Gamma(p_1)^{-\frac{1}{2}} \\
    0 \hspace{-1mm} & \hspace{-2mm} -\frac{1}{4} \Gamma(p_2)^{-\frac{3}{2}} t \hspace{-3mm} \hspace{-0.5mm} & \cdots \hspace{-1mm} & \hspace{-1mm} 0 \hspace{-1mm} & \hspace{-1mm} \frac{1}{2} \Gamma(p_2)^{-\frac{1}{2}} \\
    \vdots \hspace{-1mm} & \hspace{-1mm} \vdots \hspace{-0.5mm} & \ddots \hspace{-1mm} & \hspace{-1mm} 0 \hspace{-1mm} & \hspace{-1mm} \vdots \\
    0 \hspace{-1mm} & \hspace{-1mm} 0 \hspace{-0.5mm} & 0 \hspace{-1mm} & \hspace{-1mm} -\frac{1}{4} \Gamma(p_M)^{-\frac{3}{2}} t \hspace{-1mm} & \hspace{-1mm} \frac{1}{2} \Gamma(p_M)^{-\frac{1}{2}} \\
    \frac{1}{2} \Gamma(p_1)^{-\frac{1}{2}} \hspace{-1mm} & \hspace{-1mm} \frac{1}{2} \Gamma(p_2)^{-\frac{1}{2}} \hspace{-0.5mm} & \cdots \hspace{-1mm} & \hspace{-1mm} \frac{1}{2} \Gamma(p_M)^{-\frac{1}{2}} \hspace{-1mm} & \hspace{-1mm} 0 \\
  \end{array}
\right)
\end{equation}
where
\begin{equation}\nonumber
\Gamma(p_m)  =  \frac{P_{\max ,m}}{\eta_{\max ,m}^2} \cdot p_m \geq 0.
\end{equation}

It is clear that the Hessian matrix is neither positive nor negative, based on which~\eqref{G_EE_Opt} cannot be resolved by existing standard convex optimization algorithm directly. Therefore, the necessary optimal conditions are introduced in this paper to overcome this problem.

\vspace{2mm}

\newtheorem{theorem}{\textbf{Theorem}}
\begin{theorem} \label{PowerA_TPA}
\textit{Under TPA, the minimal energy consumption is achieved when only one BS is accessed, conveniently selected from $\mathcal{M}$. The remaining BSs are forced to stay in idle mode. Assuming that BS $m^*$ is accessed, then optimal transmit power for~\eqref{G_EE_Opt} can be obtained, as given by}
\begin{equation}\label{Pi_TPA}
    p_{m^*} = \bigg( \frac{I_{\rm out} + P_{\rm N}}{{{{\left| {{h_{m^*}}} \right|}^2}}} \bigg)  \Big( {{{\rm{2}}^{\frac{T}{W}\frac{r_{\rm dl}}{t}}} - 1} \Big),
\end{equation}
\begin{equation}\label{Pi_TPA_other}
    p_m = 0 ,\ m \ne m^*.
\end{equation}
\end{theorem}

\vspace{2mm}

\begin{IEEEproof}
Given $\mathcal{M}$ and transmit duration~$t$, an equivalent objective function can be given by
\begin{equation}\label{Opt_TPA_Fixedt}
    E_{\rm tpa}^{L} (\{ p_m \} ) = \sum_{m = 1}^M \bigg( \frac{\sqrt {{P_{\max ,m}}}}{{{\eta _{\max ,m}}}} \bigg) \sqrt {{p_m}} \, t,
\end{equation}
where we remove the constant terms of~\eqref{EE_Opt_TPA} that do not depend on transmit powers $\{p_m \}$. Then the proof will be completed in two cases, i.e., \mbox{long-term} CSI and \mbox{short-term} CSI, which are presented in detail as follows.

\vspace{2mm}

\textbf{\mbox{Long-term} CSI}: It is clear that minimizing $E_{\rm tpa}^{L}(\{ p_m \} )^2$ is equivalent to minimizing $E_{\rm tpa}^{L}(\{ p_m \} )$, due to the monotonicity of square function when $E_{\rm tpa}^{L}(\{ p_m \} ) \geq 0$. With this method, $E_{\rm tpa}^{L}(\{ p_m \} )^2$ can be expressed as
\begin{equation}\label{Opt_TPA_Fixedt2}
    E_{\rm tpa}^{L}(\{ p_m \} ) ^2 = E_{\rm tpa}^{\rm eq}(\{ p_m \} ) + E_{\rm}^{\rm plus}(\{ p_m \} ),
\end{equation}
where
\begin{eqnarray}
E_{\rm tpa}^{\rm eq}(\{ p_m \} ) \hspace{-2mm} &=& \hspace{-2mm} \sum_{m = 1}^M \bigg( \frac{{{P_{\max ,m}}}}{{\eta _{\max ,m}^2}} \bigg) \, p_m \, t^2 , \label{Opt_TPA_eq}\\
E_{\rm tpa}^{\rm plus}(\{ p_m \} ) \hspace{-2mm} &=& \hspace{-2mm} \sum_{m \ne n} {\frac{{\sqrt {{P_{\max ,m}}{P_{\max ,n}}} }}{{{\eta _{\max ,m}}{\eta _{\max ,n}}}} \, \sqrt{{p_m} \, {p_n}} \, {t^2}} \label{Opt_TPA_plus}.
\end{eqnarray}
In the following, we first prove that the transmit power presented in the theorem above minimizes~\eqref{Opt_TPA_eq} and then, prove that this solution minimizes~\eqref{Opt_TPA_Fixedt2} as well.

The Lagrange function that combines the objective function and data rate constraint of this new optimization problem can be formulated, as given by
\begin{equation}
    {\mathcal{L}_{\rm{tpa},1}} = E_{\rm tpa}^{\rm eq} - \lambda_{{\rm tpa},1} \bigg( \frac{t}{T} \bigg)  W  \log_2 \bigg( 1 + \frac{S}{I_{\rm out} + P_{\rm N}} \bigg),
\end{equation}
where $\lambda_{{\rm tpa},1}$ is the Lagrange multiplier. According to \mbox{Karush-Kuhn-Tucker}~(KKT) conditions, \mbox{$\frac{{\partial {\mathcal{L}_{{\rm tps},1}}}}{{\partial {p_m}}} = 0$} should hold simultaneously for all BSs from $\mathcal{M}$ in any local optimal solution. Therefore, considering $0 \le t \le T$, the following relationship must be verified for all $m \in \mathcal{M}$, as expressed by
\begin{equation}\label{KKT_Con_TPA}
  \bigg(  \frac{{\left| {{h_m}} \right|}^2}{I_{\rm out} + P_{\rm N}} \bigg)  \bigg( \frac{\eta _{\max ,m}^2}{P_{\max ,m}} \bigg) = \bigg( \frac{T \log_e 2}{{\lambda_{{\rm tpa},1}} W } \bigg) \bigg( 2^{\frac{T}{W}\frac{r_{\rm dl}}{t}} \bigg) \, t .
\end{equation}

By exploring the structure of~\eqref{KKT_Con_TPA}, it is observed that only one BS can be accessed by the typical UE in any local optimal solution. This is because the \mbox{right-hand} side~(RHS) of~\eqref{KKT_Con_TPA} is constant regardless of $m$, whereas the \mbox{left-hand} side~(LHS) takes a different values for each different BS index. Therefore, a different local optimal solution is obtained by first accessing a given BS~$m^*$, then selecting its corresponding transmit power~\eqref{Pi_TPA} with the aid of~\eqref{AchiRate}, and finally forcing $p_m = 0$ for $m \ne m^*$. In this case, when evaluating~\eqref{Opt_TPA_plus}, $E_{\rm tpa}^{\rm plus}(\{ p_m \} ) = 0$, which is its minimal value, and in turn~\eqref{Opt_TPA_Fixedt} reaches its minimal value. Moreover, the second term in~\eqref{EE_Opt_TPA} is also minimal if only one BS is accessed. Therefore, the proposed solution represents a local optimum for~\eqref{Opt_TPA_plus}.

\vspace{2mm}

\textbf{\mbox{Short-term} CSI}: In this case, we directly minimize $E_{\rm tpa}^{L}(\{ p_m \} )$ in \eqref{Opt_TPA_Fixedt}, where the Lagrange function for $E_{\rm tpa}^{L}(\{ p_m \} )$ can be formulated, as given by
\begin{equation}
    \mathcal{L}_{{\rm tpa},2} = E_{\rm tpa}^{L} - \lambda_{{\rm tpa},2}  \bigg( \frac{t}{T} \bigg)  W  \log_2\bigg( 1 + \frac{S}{ I_{\rm out} + P_{\rm N} } \bigg).
\end{equation}

Similarly, according to the KKT conditions, any local optimal solution for $0 \leq t \leq T$ should verify $\frac{{\partial {\mathcal{L}_{{\rm tpa},2}}}}{{\partial {p_m}}} = 0$ simultaneously for all BSs from $\mathcal{M}$, which can be transformed, as expressed by
\begin{equation}\label{KKT_Con_TPA_Coh}
\bigg( \frac{|h_m|}{\sqrt{I_{\rm out} \hspace{-0.75mm} + \hspace{-0.75mm} P_{\rm N}}} \bigg) \hspace{-0.75mm} \bigg( \frac{\eta_{\max,m}}{{\sqrt {{P_{\max ,m}}} }} \bigg) \hspace{-0.5mm} = \hspace{-0.5mm} \bigg( \frac{T \log_e 2}{2 \lambda_{{\rm tpa},2} W} \bigg) \hspace{-0.75mm} \Bigg( \frac{2^{\frac{T}{W}\frac{r_{\rm dl}}{t}}}{\sqrt{{2^{\frac{T}{W}\frac{r_{\rm dl}}{t}}} \hspace{-0.75mm} - \hspace{-0.75mm} 1 } } \Bigg) \hspace{-0.25mm} . \hspace{-0.5mm}
\end{equation}

Like the \mbox{long-term} CSI case, since the RHS of~\eqref{KKT_Con_TPA_Coh} is constant regardless of $m$, whereas the LHS takes a different values for each different BS index, it can be shown that only one BS is accessed for any local optimum, and its transmit power can be calulated with the help of~\eqref{Pi_TPA} to minimize the total energy consumption $E_{\rm tpa}$.

Combining the two different cases, the theorem is completely proved.
\end{IEEEproof}

\vspace{2mm}

\textbf{Remark:} Unlike maximizing the SE in HCNs, when considering the minimization of total energy consumption, only one BS is accessed at any scheduling time interval in both \mbox{long-term} and \mbox{short-term} CSI cases under non-ideal circuitry.

\vspace{2mm}

Let us now focus on the optimization of the transmit duration $t$ when the energy consumption of BS $m$ is evaluated. Combining~\eqref{Pi_TPA} with \eqref{EE_Opt_TPA}, the transmit duration optimization problem can be reformulated as
\begin{equation}\label{Re_Opt_TPA}\tag{\textbf{P2}}
\begin{aligned}
{\min\limits_t} \hspace{-0.5mm} \ E_{\rm tpa}^{m}(t) \hspace{-1mm} &=  \hspace{-1mm} \bigg( \hspace{-0.5mm} \frac{\sqrt{{P_{\max ,m}}}}{\eta _{\max,m}} \bigg) \hspace{-0.5mm} \Bigg[ \hspace{-0.5mm} \sqrt{ \bigg( \hspace{-0.5mm} \frac{ I_{\rm out} \hspace{-0.75mm} + \hspace{-0.75mm} P_{\rm N} }{\left| {{h_m}} \right|^2} \bigg) \hspace{-0.75mm} \bigg( \hspace{-0.5mm} {{2^{\frac{T}{W}\frac{r_{\rm dl}}{t}}} \hspace{-0.75mm} - \hspace{-0.75mm} 1} \bigg)} \, \Bigg]  t  \\
&+ \left(\varepsilon_m r_{\rm dl} + P_{{\rm base},m} - P_{{\rm idle},m}\right) t  \\
&+ \left(\varepsilon_{\rm u} r_{\rm dl} + P_{{\rm base}, {\rm u}} - P_{{\rm idle},\rm {u}}\right) t
\end{aligned}
\end{equation}
\begin{equation}\nonumber
{\rm s.t.}\ \ 0 \le t \le T.
\end{equation}

It can be easily proved that energy consumption minimization problem~\eqref{Re_Opt_TPA} is convex when \mbox{$\frac{r_{\rm dl}}{W}>1$}, since the \mbox{second-order} derivative of the objective function in this situation is always positive.

For minimizing the total energy consumption, the selection criterion of optimal BS can be written as
\begin{equation}\label{SelectCriterion_TPA}
m^*_{\rm tpa} = \arg\min_m \, E_{\rm tpa}^{m}(t).
\end{equation}

To reduce the complexity and obtain closed-form expressions, two approximate relationships can be given by
\begin{equation}\nonumber
{{\rm{2}}^{\frac{T}{W}\frac{r_{dl}}{t}}} \gg 1,
\end{equation}
\begin{equation}\nonumber
\Big({\rm{2}}^{\frac{T}{W}\frac{r_{\rm dl}}{t}} - 1 \Big) \approx \Big( {\rm{2}}^{\frac{T}{W}\frac{r_{\rm dl}}{t}} \Big),
\end{equation}
which can both be established in low and high data rate region. If the system works in high data rate region, the SE $\frac{{r_{\rm dl}}}{W}$ is large enough so that ${{\rm{2}}^{\frac{T}{W}\frac{r_{dl}}{t}}} \gg 1$. This is the typical case nowadays, where the advanced technologies, e.g., high-order modulation, are exploited to increase the SE. And in low data rate region, the transmit duration will be reduced to be small enough to weaken the effect of the circuit power for achieving minimal energy consumption, based on which $\frac{T}{t}$ is very large so that ${{\rm{2}}^{\frac{T}{W}\frac{r_{dl}}{t}}} \gg 1$~\cite{Refenece19}.

It is noted that the transmit power is very large enough to dominate the overall power consumption in the approximate situation so that the circuit power can be ignored. Substituting the approximate relationships into the objective function of~\eqref{Re_Opt_TPA}, a new approximate optimization can be reformulated. Then, it can be shown that the \mbox{first-order} derivative of the approximate objective function of~\eqref{Re_Opt_TPA} equals zero when following conditions are verified, i.e., the optimal transmit duration $t^*_{\rm tpa}$ can be formulated, as given by
\begin{equation}\label{t_TPA}
    t^*_{\rm tpa} = \left\{
		\begin{array}{ll}
    t_{\min}, & \text{if  }  \big( \frac{{r_{\rm dl}T\ln 2}}{{2W}} \big) < {t_{\min }},\\
    T  ,       & \text{if  }  \big( \frac{r_{\rm dl}}{W} \big) > \big( \frac{2}{{\log_e 2}} \big), \\
    \frac{{r_{\rm dl}T\ln 2}}{{2W}}, & \text{otherwise}.
    \end{array} \right.
\end{equation}

And if the circuit power can be ignored, the selection criterion for the optimal BS can be simplified, as given by
\begin{equation}\label{ApprSC_TPA}
    m^*_{\rm tpa} = \arg \max_m \bigg( \frac{{{\eta_{\max,m}} \, {{\left| {{h_m}} \right|}}}}{{\sqrt {{P_{\max ,m}}} }} \bigg).
\end{equation}
\section{Algorithmic Implementation}\label{sec:4}
With the aid of Theorem~\ref{PowerA_TPA}, the \mbox{non-convex} energy consumption minimization problem can be reformulated into a set of convex optimization problems with the structure of~\eqref{Re_Opt_TPA}. These equivalent optimization problems can be solved with the aid of standard convex optimization algorithms, like the linear search method in this paper. Moreover, in the approximate situation, the optimal RA can simply be obtained through approximate criterion and expressions. A summary of this procedure is summarized in Algorithm~\ref{alg1}.
\begin{algorithm}[!t]
\caption{The proposed scheme}
\label{alg1}
\begin{algorithmic}[1]
\STATE Feedback all estimated channel gain $h_m$ from $\mathcal{M}$; \\
\vspace{2mm}
\hspace{-7mm} \textbf{Case (a): Precise situation}
\FOR{$m = 1 \text{ to } M$}
    \STATE Run linear search to get optimal $t_m$ from problem \textbf{(P2)};
    \STATE Compute $p_m$ using \eqref{Pi_TPA} and compute $E_{\rm tpa}^m$;
\ENDFOR
\STATE Access BS $m_{\rm tpa}^*$ using \eqref{SelectCriterion_TPA} and record $p_{m^*}$ and $t_{\rm tpa}^*$; \\
\vspace{2mm}
\hspace{-7mm} \textbf{Case (b): Approximate situation}
\STATE Access BS $m_{\rm tpa}^*$ with the aid of \eqref{ApprSC_TPA};
\STATE Find (and record) $t_{\rm tpa}^*$ and $p_{m^*}$ using \eqref{t_TPA} and \eqref{Pi_TPA};
\vspace{1mm}
\STATE Inform $p_{m^*}$ and $t_{\rm tpa}^*$ to BS $m_{\rm tpa}^*$, while other BSs are in idle mode;
\end{algorithmic}
\end{algorithm}
\section{Simulation results}\label{sec:5}
Simulations are carried out to validate our proposed scheme in HCNs with non-ideal circuitry. Apart from the proposed algorithm, some state-of-the-art approaches are also simulated for comparison purpose, where traditional scheme is to access single serving BS that can provide maximum received signal strength~\cite{Refenece13}, while all access scheme is to access all BSs with uniformly allocated transmit powers~\cite{Refenece20}. In is noted that all these simulated schemes are with optimized transmit duration, unless otherwise specified~\cite{Refenece18,Refenece19}. Moreover, the \mbox{long-term} and \mbox{short-term} CSI cases degenerate to the identical results and attain the same performance consequently; see Theorem~\ref{PowerA_TPA}. Therefore, we do not distinguish the two cases and plot them by one curve in the simulations.

For fair comparison, all the simulated schemes have the same required data rate $r_{\rm dl}$ with time frame $T$. The circuit power ranges from tens to hundreds of milliWatts, comprised mainly of baseband processing and radio-frequency~(RF) generation. The reasonable range of PA efficiency at the maximum average BS transmit power is around $0.311$ to $0.388$. The simulation parameters are listed in Table~\ref{tab:1}.

\begin{table}[t]
\vspace{-3mm}
\centering
\caption{Simulation Parameters}
\vspace{-2mm}
\begin{tabular}{|l|l|}
\hline
\textbf{Parameters} & \textbf{Values} \\
\hline
System bandwidth ($W$) & 10\,[MHz] \\
Time frame duration ($T$) & 10\,[ms] \\
Noise power spectral density ($N_0$) & --\,174\,[dBm/Hz] \\
The density of BSs & 20\, [BS/$\rm km^2$] \\
Length of research area & 300\,[m] \\
Mean path loss model ($L_{m,{\rm u}}$) & $103.8+21\log_{10}(d)$ \,[dB] \\
Idle power consumption ($P_{{\rm idle},{\rm m}}$, $P_{{\rm idle},{\rm u}}$) & 30, 10\,[mW] \\
Static circuit power ($P_{{\rm base},{\rm m}}$, $P_{{\rm base},{\rm u}}$) & 50, 20\,[mW] \\
Dynamic circuit factor ($\varepsilon_{{\rm m}}$, $\varepsilon_{{\rm u}}$) & 5, 2\,[mW/Mbps] \\
Maximum output power ($P_{\max,m}$) & 46\,[dBm] \\
Maximum PA efficiency ($\eta_{\max,m}$) & 0.35\\
\hline
\end{tabular}
\label{tab:1}
\vspace{-3mm}
\end{table}

\begin{figure}[!t]
\centering
\vspace{-2mm}
\includegraphics[scale=0.28]{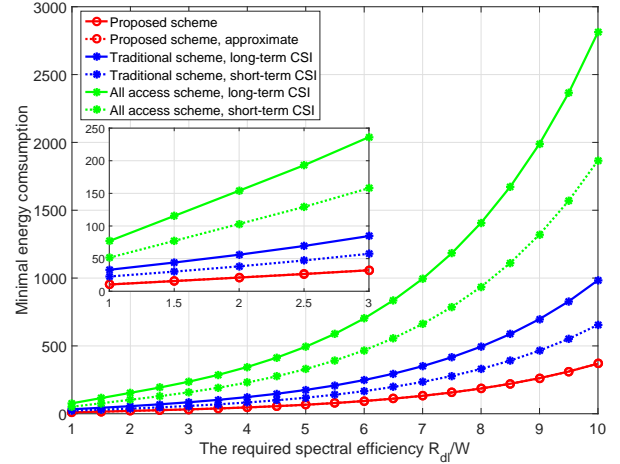}
\vspace{-8mm}
\caption{The minimal energy consumption as function of required SE considering both TPA and circuit power.}
\label{EEPerfprmance}
\vspace{-4mm}
\end{figure}
\begin{figure}[!t]
\centering
\vspace{-2mm}
\includegraphics[scale=0.28]{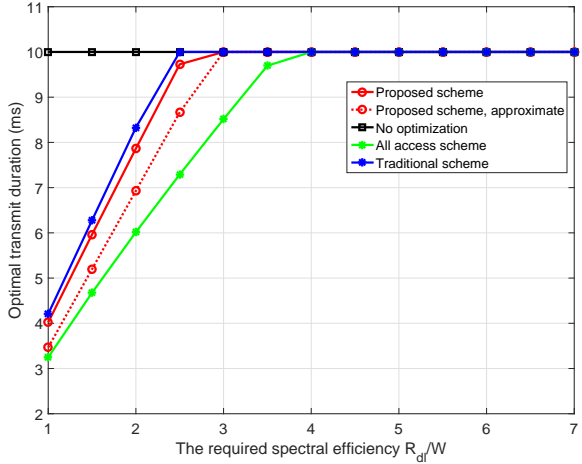}
\vspace{-8mm}
\caption{The optimal transmit duration as function of required SE for different schemes under non-ideal circuitry.}
\label{TimeDuration}
\vspace{-4mm}
\end{figure}

Fig.~\ref{EEPerfprmance} shows the minimal energy consumption for different scheme under non-ideal circuitry. It is obvious that our proposed scheme outperforms the others compared schemes in any case. Although the all access scheme can get the maximal system capacity, its energy performance is the worst due to quite large power consumed for non-ideal circuitry at all BSs. It is seen that the minimal energy consumption of the proposed scheme exhibits a linear growth in the low required SE region and, after a threshold, the energy starts to increase exponentially. This is because with low required SE, the circuit power dominates the total energy consumption and the transmit duration will increase to support the rising required data rate, based on which the total energy consumption grows linearly corresponding to the linear increase of transmit duration, as shown in Fig.~\ref{TimeDuration}. However, when the required SE grows, the transmit duration starts to occupy most part of the frame and the energy consumption starts to increase exponentially because transmit power needs to grow exponentially to maintain the target data rate. Under TPA, both cases of \mbox{long-term} and \mbox{short-term} CSI provide the same energy performance for proposed scheme, whereas the \mbox{short-term} CSI case gives a better energy performance for other schemes due to the larger received signal power with knowledge of phase information. Finally, it is observed that the two curves of precise and approximate cases are almost overlapped, which indicates the difference between the precise and approximate expressions for minimal energy consumption is negligible.

Fig.~\ref{TimeDuration} plots the optimal transmit duration for different schemes with non-ideal circuitry. It is possible to observe that our proposed duration optimization can leverage the transmit power and duration optimally. It is seen that when the required SE is less than $4$ bps/Hz, the transmit duration is less than the entire time and grows linearly with the increase of SE requirement. It aims to avoid the exponential increase of transmit power and the extra power consumed in TPA. However, when required SE is over $4$ bps/Hz, the entire frame is occupied for transmission, and the optimal transmit power need to increase exponentially to meet the increasing required SE. It also can be seen that the optimal transmit duration of all access scheme is shorter than the other counterparts, because it contains much more circuit power which needs less transmit duration to minimize this effect. Moveover, since the traditional scheme cannot access the most energy-efficient BS, its transmit duration is a little larger than the proposed scheme to reduce the transmit power. Finally, it is possible to see that approximate solution~\eqref{t_TPA} of problem~\eqref{Re_Opt_TPA} is tight to some extent, even in the low required SE range.

\begin{figure}[!t]
\centering
\vspace{-2mm}
\includegraphics[scale=0.28]{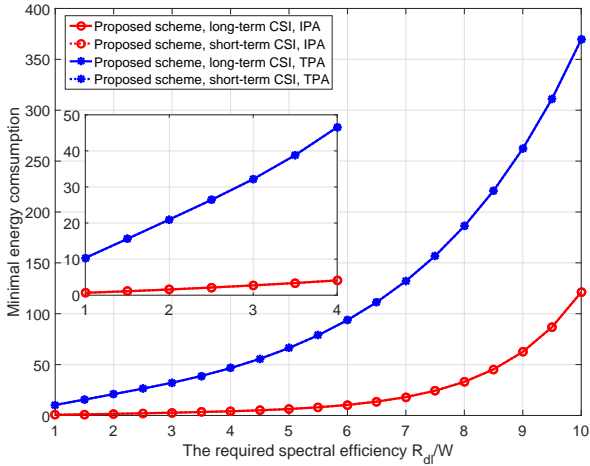}
\vspace{-8mm}
\caption{The minimal energy consumption as function of required SE for different schemes using either TPA or IPA model with circuit power.}
\label{EECompare}
\vspace{-4mm}
\end{figure}
Fig.~\ref{EECompare} compares the minimal energy consumption of both TPA and IPA. The minimal energy consumption is lower for IPA than TPA, particularly in the low required SE region. This is because the power consumption of TPA is higher than IPA for the same transmit power. When compared to IPA, the curves for TPA do not exhibit the flat characteristic, which means that the consideration of \mbox{non-ideal} PAs increases the total energy consumption significantly for the same required SE. This is the same reason why the linear growth rate of TPA curve is much bigger than that of IPA curve in the low required SE region. It also can be seen that the cases of \mbox{long-term} and \mbox{short-term} CSI can get the same energy performance, which verifies the analyzed results of Theorem~\ref{PowerA_TPA}.
\section{Conclusion}\label{sec:6}
In this paper, we investigated the energy aspect of UAC based on optimal RA in HCNs, with the consideration of both the \mbox{non-ideal} PAs and circuit power. Through the optimal RA, BS access and transmit duration optimization are derived. It is proved that single BS access is the best scheme and, making use of this result, the original \mbox{multi-variable} non-convex optimization problem was transformed into a series of \mbox{single-variable} convex problems that were solved using the linear search method. Numerical simulations were carried out to validate the analysis, and to characterize the effect of \mbox{non-ideal} circuitry on energy performance of HCNs.
\section*{Acknowledgement}
The work was supported in part by National Nature Science Foundation of China Project under Grant 61471058, in part by the Key National Science Foundation of China under Grant 61461136002, in part by the Hong Kong, Macao and Taiwan Science and Technology Cooperation Projects under Grant 2016YFE0122900, and in part by the 111 Project of China under Grant B16006​.

\end{document}